\documentclass[usenatbib,fleqn,useAMS]{mnras}

 \usepackage{graphicx}
 \usepackage{amsmath}
 \usepackage{mathrsfs}
 \usepackage[varg]{txfonts}
 \usepackage{braket}
 \usepackage{cancel}
 \usepackage{color}

 \usepackage{soul}   

 \renewcommand\le\oldleq
 \renewcommand\ge\oldgeq

\newcommand{\msun}{M_\odot}

 \renewcommand\pi\upi

 \title[Streams and Clumps]
 {Halo Substructure in the SDSS-Gaia Catalogue: Streams and Clumps}

 \author[Myeong et al.]
        {G.~C.~Myeong$^1$\thanks{E-mail:~gm564,nwe,vasilyast.cam.ac.uk,
            koposov@cmu.edu, nicola.amorisco@cfa.harvard.edu},
          N.~W.~Evans$^1$, V.~Belokurov$^1$, N.C. Amorisco$^{2,3}$ \&
          S.E.~Koposov$^{1,4}$ \\$^1$Institute of Astronomy,
          University of Cambridge, Madingley Road, Cambridge CB3~0HA
          \\$^2$ Max Planck Institute for Astrophysics, Karl
          Schwarzschild Strasse 1, D-85740 Garching, Germany \\$^3$
          Institute for Theory and Computation, Harvard Smithsonian
          Center for Astrophysics, 60 Garden Street, Cambridge,
          MA02138, USA \\$^4$ McWilliams Center for Cosmology,
          Department of Physics, Carnegie Mellon University, 5000
          Forbes Avenue, Pittsburgh, PA 15213, USA}

 \pagerange{\pageref{firstpage}--\pageref{lastpage}}\volume{000}\pubyear{2017}
 \date{version \today.}

\begin{document}
 \label{firstpage}
 \maketitle

 \begin{abstract}
 We use the SDSS-Gaia Catalogue to identify six new pieces of halo
 substructure. SDSS-Gaia is an astrometric catalogue that exploits
 SDSS data release 9 to provide first epoch photometry for objects in
 the Gaia source catalogue. We use a version of the catalogue
 containing $245\,316$ stars with all phase space coordinates within a
 heliocentric distance of $\sim 10$ kpc. We devise a method to assess
 the significance of halo substructures based on their clustering in
 velocity space. The two most substantial structures are multiple
 wraps of a stream which has undergone considerable phase mixing (S1,
 with 94 members) and a kinematically cold stream (S2, with 61
 members). The member stars of S1 have a median position of ($X,Y,Z$)
 = ($8.12, -0.22, 2.75$) kpc and a median metallicity of [Fe/H] $=
 -1.78$. The stars of S2 have median coordinates ($X,Y,Z$) = ($8.66,
 0.30, 0.77$) kpc and a median metallicity of [Fe/H] $= -1.91$. They
 lie in velocity space close to some of the stars in the stream
 reported by Helmi et al. (1999). By modelling, we estimate that both
 structures had progenitors with virial masses $\approx 10^{10} \msun$
 and infall times $\gtrsim 9$ Gyr ago. Using abundance matching, these
 correspond to stellar masses between $10^6$ and $10^7
 \msun$. These are somewhat larger than the masses inferred through
 the mass-metallicity relation by factors of 5 to 15. Additionally, we
 identify two further substructures (S3 and S4 with 55 and 40 members)
 and two clusters or moving groups (C1 and C2 with 24 and 12)
 members. In all 6 cases, clustering in kinematics is found to
 correspond to clustering in both configuration space and metallicity,
 adding credence to the reliability of our detections.
 \end{abstract}

 \section{Introduction}

 Lord Rutherford briskly asserted ``All Science is either Physics or
 Stamp Collecting''. The study of the stellar halo of the Milky Way
 has seen much philately over the last decade with the discovery of
 abundant streams and substructure~\citep[e.g.,][]{Be06,Gr09,Ne15}.
 These have usually been identified as overdensities from resolved
 star maps. Substructures remain kinematically cold and identifiable
 in phase space long after they have ceased to be recognizable in star
 counts against the stellar background of the Galaxy. In principle,
 searches in velocity space or in phase space are much more powerful
 than direct searches in configuration space. There are believed to be
 hundreds of accreted dwarf galaxies and globular clusters in the halo
 of the Milky Way which could be found through searches in velocity
 space.

 In practice, kinematic data has been so fragmentary to date that such
 substructure searches have been difficult to perform. There have
 been some successes, such as the group of 8 stars in the Hipparcos
 data clumped in metallicity and phase space found by \citet{He99} or
 the discrete kinematic overdensities in Sloan Digital Sky Survey
 (SDSS) Stripe 82 identified by \citet{Sm09}. Nonetheless, given the
 ostensible power of the method, results have been meagre.

 The advent of data from the Gaia satellite~\citep{Gaia1} is a pivotal
 moment for identifying the hundreds of partially mixed phase space
 structures that numerical simulations suggest should be present in the
 halo. Many of these have dissolved sufficiently to fall below the
 surface brightness threshold of current imaging surveys, and thus
 will remain unnoticed without kinematic data from Gaia.

 The first Gaia data release provided TGAS, or the Tycho-Gaia
 Astrometric Solution, which used the earlier Tycho catalogue as the
 first epoch for the astrometric solution~\citep{Gaia2}. TGAS gives
 the proper motions and parallaxes of just over 2 million
 stars. Subsets of these stars are in ongoing radial velocity surveys
 such as LAMOST, RAVE or RAVE-on~\citep{Lu15,Ca17,Ku17}.  Already,
 claims of a coherently moving feature in velocity space ~\citep{My17}
 as well as over-densities in ``integrals of motion
 space''~\citep{He17} have been made.

 Cross-matches between TGAS and radial velocity surveys produce
 catalogues of $\sim$ 250\,000 stars. These are primarily local
 samples, dominated by denizens of the local disk within 1 kpc.  It
 would be advantageous to use a much larger and deeper sample of stars
 with full phase space information. Along with TGAS, Gaia data release
 1 also comprised the Gaia source photometric catalogue, which
 provides the locations of $\sim 10^9$ sources. Koposov (2017, in
 prep.) recalibrated the SDSS astrometric solution and then obtained
 proper motions from Gaia positions and their recalibarted positions
 in SDSS. This catalogue is also discussed in some detail in
 \citet{De17} and \citet{DeBoer18}. The individual SDSS-Gaia proper
 motions have statistical errors typically $\sim 2$ mas yr$^{-1}$, or
 $\sim 9.48 D$ km s$^{-1}$ for a star with heliocentric distance $D$
 kpc. As the SDSS data were taken over a significant period of time,
 the error is primarily controlled by the time baseline. However,
 there are no systematic effects down to a level of $0.1 - 0.2$ mas
 yr$^{-1}$ in the astrometry with regard to magnitude or colour
 \citep[see e.g., Figure 2 of ][]{De17}, so this makes the SDSS-Gaia
 catalogue suitable for searching for large-scale velocity signatures
 corresponding to streams and substructures.

 The depth of SDSS-Gaia enables us to search out to heliocentric
 distances of $\sim 10$ kpc, which is a substantial advantage over
 TGAS. Cross-matching SDSS-Gaia with spectroscopic surveys can add
 radial velocities. Finally, photometric parallaxes for stars such as
 main-sequence turn-offs (MSTOs) or blue horizontal branch stars
 (BHBs) gives samples with the full six-dimensional phase space
 coordinates. Although the SDSS-Gaia catalogue will be superseded in
 April 2018 by the next Gaia data release, it currently provides the
 best catalogue in which to search for halo substructure by kinematic
 means.

 The overall aim of this activity is to constrain the fraction of halo
 stars in clumps and substructures. This is of great interest as it
 encodes the accretion history of the stellar halo and by extension of
 the Milky Way itself. Nonetheless, the optimum algorithms for
 substructure identification, as well as the best methodologies to
 match detected substructures to disrupting subhalos in numerical
 simulations, are ripe for exploration with SDSS-Gaia. Ultimately, a
 better understanding of such algorithms is needed to convert the
 'stamp collecting' into astrophysics.

 In this spirit, Section 2 introduces a new method to search for
 substructure in velocity space in the SDSS-Gaia catalogue. The six
 most significant halo substructures are studied in detail in Section
 3. They include a gigantic stream with cold kinematics, two moving
 groups and three hotter substructures in which the velocity
 distribution in at least one component is very broad. By matching
 with substructure in a library of numerical simulations in
 Section 4, we argue that hotter substructures probably correspond to
 multiply-wrapped streams in the later stages of disruption. For the
 two largest substructures, we provide estimates of the likely mass of
 the progenitor and infall time.  Finally, Section 5 sums up with an
 eye to possible extensions and elaborations of our new method.

 \begin{figure*}
 \begin{center}
 \includegraphics[width=0.9\textwidth]{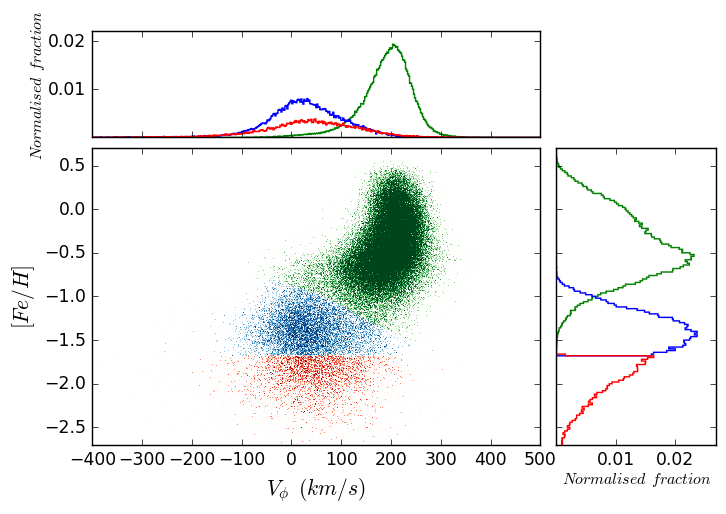}
 \end{center}
 \caption{The cleaned sample is shown in the ($v_\phi, [{\rm Fe/H}]$)
   plane. There is a clear separation of the halo stars from the disk
   (thin and thick) populations.  Green represents the disk, blue the
   relatively metal-rich halo ([Fe/H] $> -1.65$), and red the
   relatively metal-poor halo ([Fe/H] $< -1.65$). For the one
   dimensional $v_\phi$ and [Fe/H] distributions, the normalisation is
   performed separately for the disk, and for the entire halo group,
   so the sum of the area under the green histogram is unity, as is
   that for the blue and red combined.}
 \label{fig:figone}
\end{figure*}
\begin{figure*}
 \begin{center}
 \includegraphics[width=0.9\textwidth]{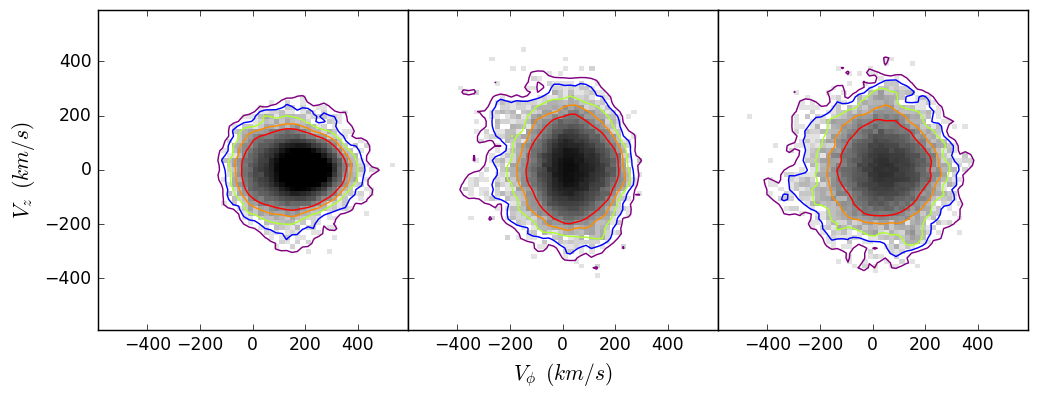}
 \end{center}
\caption{The data are shown in the plane of ($v_\phi, v_z$) for the
  disk (left), metal-rich halo (middle) and metal-poor halo
  (right). The contours levels are logarithmic. We can see visible
  substructure evident in the metal-rich ([Fe/H] $> -1.65$) and
  metal-poor ([Fe/H] $< -1.65$) halo groups. It is apparent that the
  sequence from disk to metal-rich halo to metal-poor halo is one of
  increasing lumpiness and substructure. The pixel size is 20
  kms$^{-1}$ on each side. The outermost contour is 2 stars per
  pixel}, and the contours increase by a factor of $10^{0.35} \approx
  2.24$ on moving inwards.
 \label{fig:lumpy}
 \end{figure*}
 \begin{figure*}
 \begin{center}
 \includegraphics[width=0.9\textwidth]{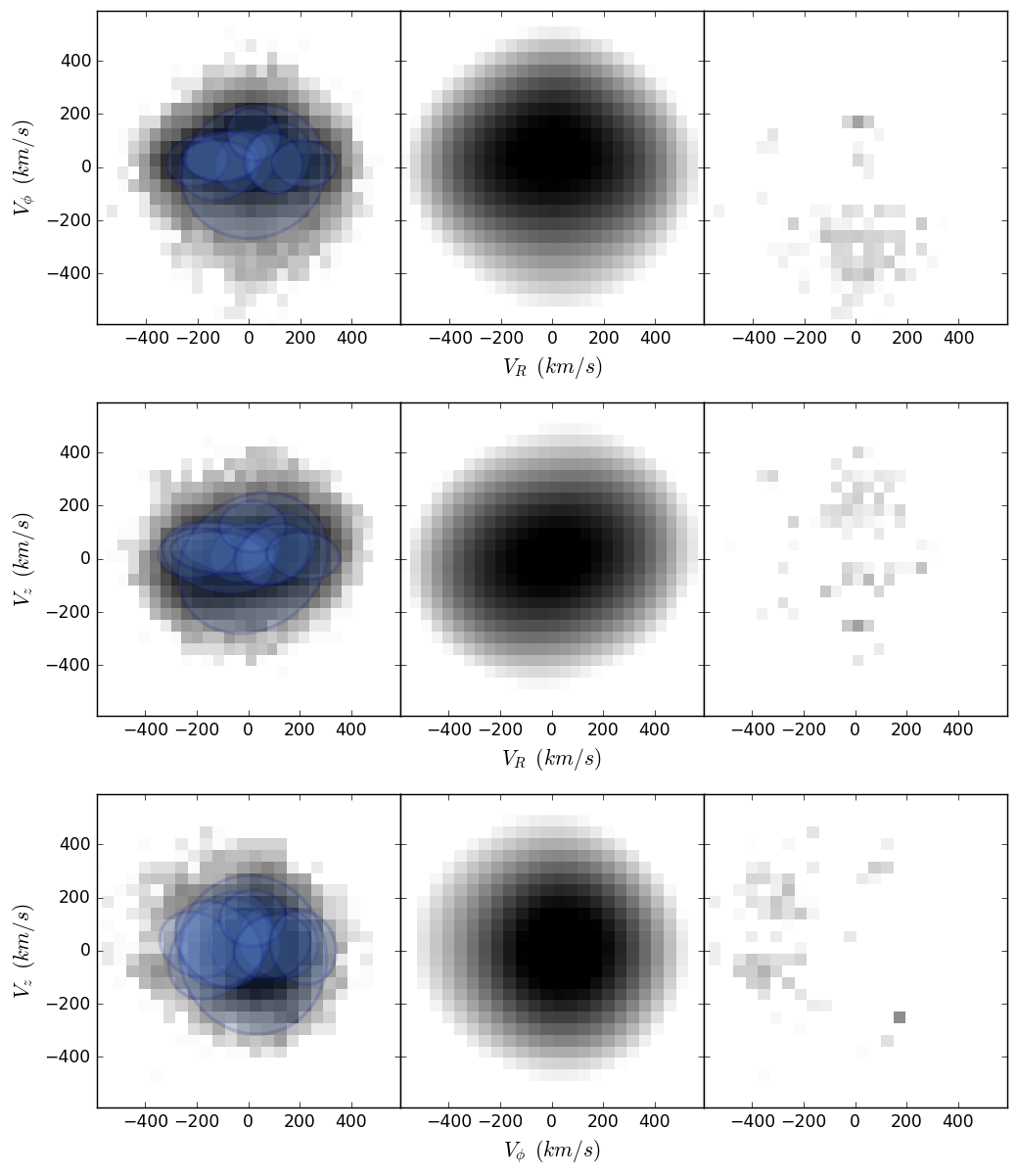}
 \end{center}
\caption{For the entire halo sample, we show from left to right the
  data, the smooth Gaussian Mixture model, and the residuals.
  Superposed on the data are blue ellipses representing the Gaussians
  with orientation and sizes scaled according to their principal
    axes.  The rows show the principal planes in velocity space
  ($v_R, v_\phi$), ($v_R, v_z$) and ($v_\phi, v_z$)
  respectively. Although the Gaussian mixture model is a good
  representation of the halo, substructure is already apparent in the
  plots in the rightmost column. The residuals demonstrate the
  locations of the main pieces of substructure, as well as
  highlighting the lumpy nature of the distribution.}
 \label{fig:residuals}
 \end{figure*}
\begin{figure*}
 \begin{center}
   \includegraphics[width=0.9\textwidth]{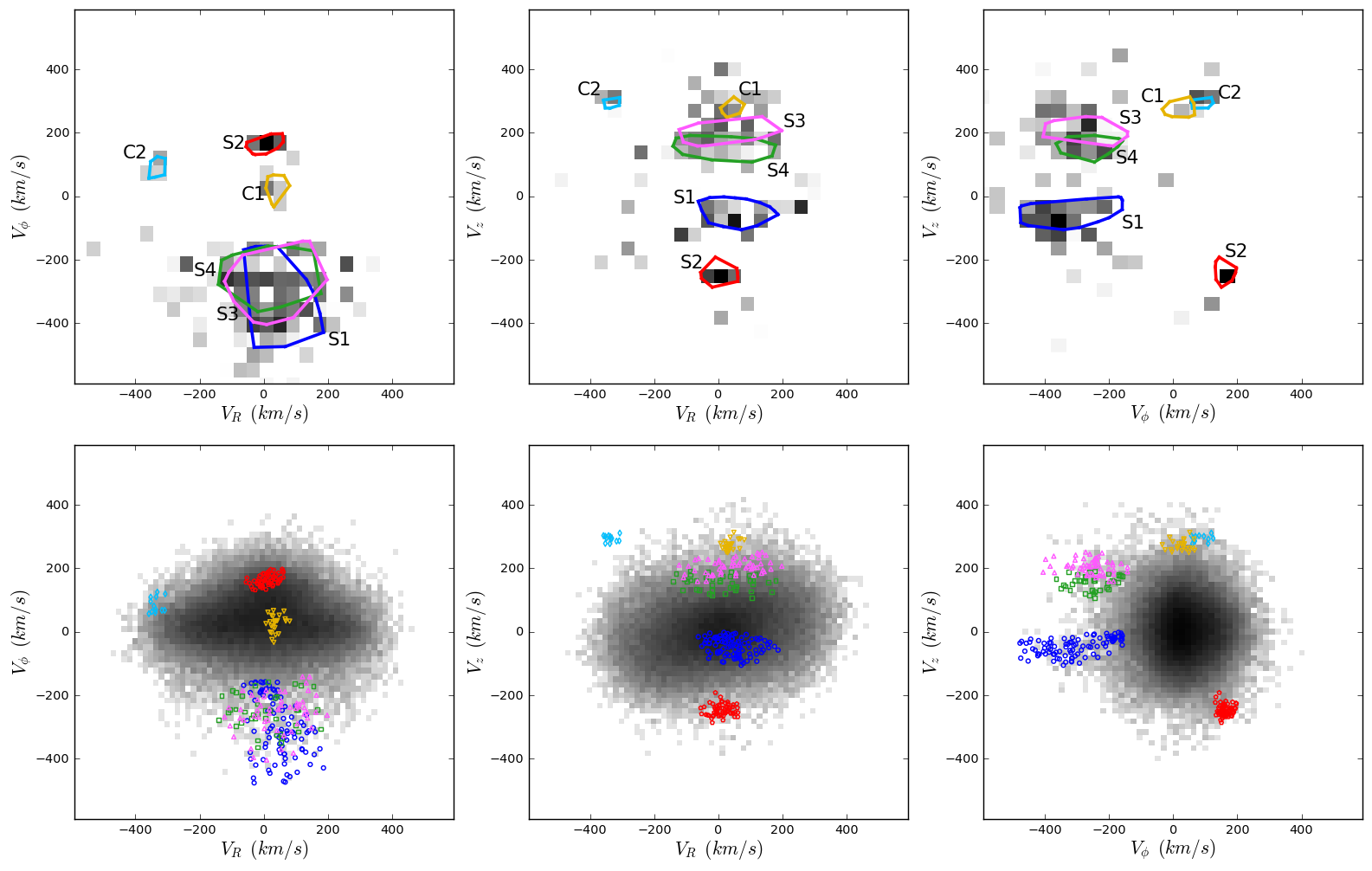}
 \end{center}
 \caption{The velocity distributions of the full halo sample (bottom
   row) and the residuals (top row) are shown in the three principal
   planes in velocity space, ($v_R,v_\phi$), ($v_R, v_z$) and
   ($v_\phi, v_z$).  Stars belonging to the two most prominent
   substructures are shown as blue circles and red pentagons (S1 and
   S2). Also shown are two smaller substructures as upward-pointing
   magenta triangles and green squares (S3 and S4), and two moving
   clumps as downward-pointing brown triangles and pale blue diamonds
   (C1 and C2).}
 \label{fig:figvd}
 \end{figure*}
 \begin{figure*}
 \begin{center}
 \includegraphics[width=0.9\textwidth]{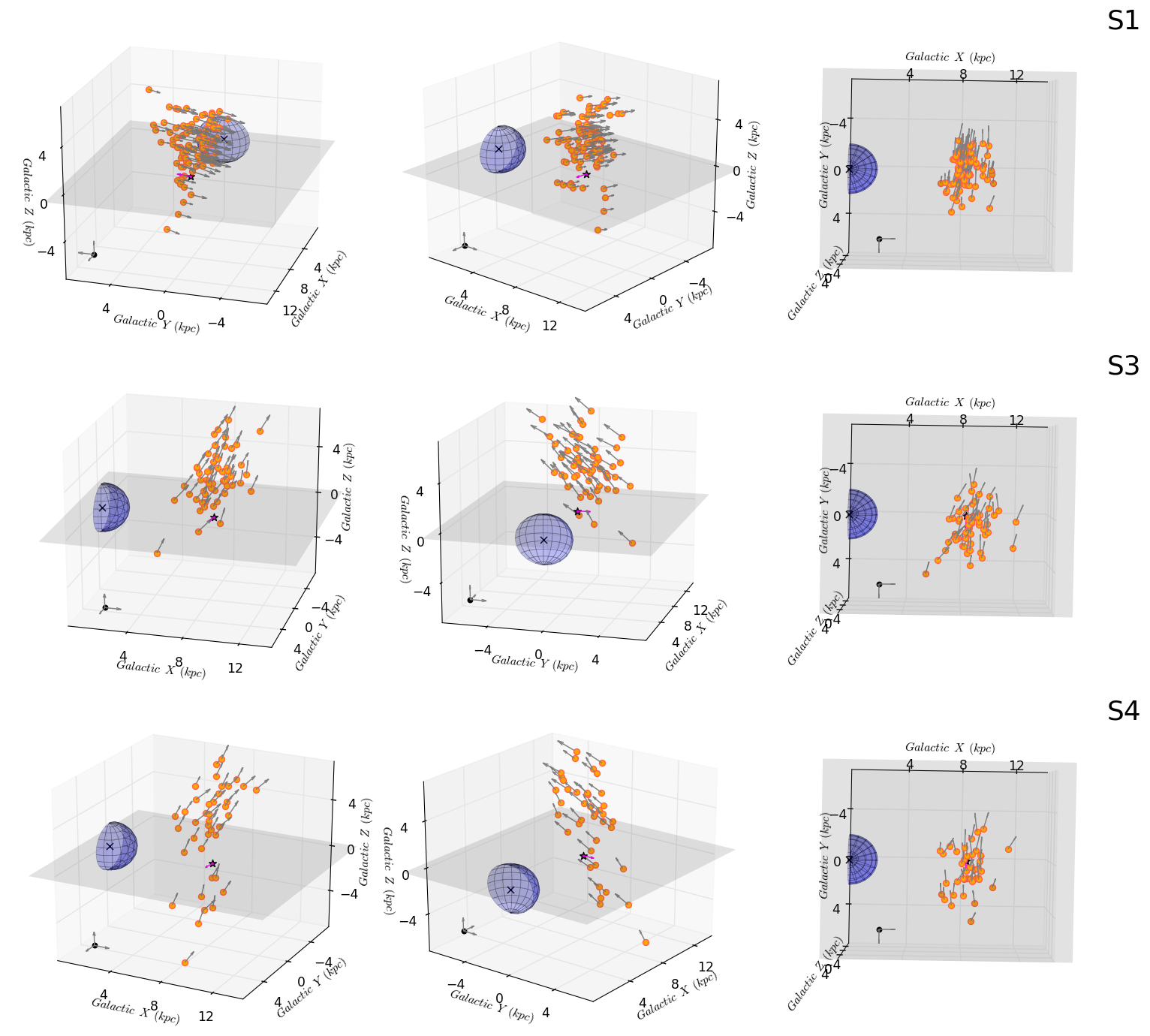}
 \end{center}
 \caption{The properties of stars belonging to the three substructures
   S1, S3 and S4. We have grouped them together because of the
   morphological similarity. The left and middle panes show two views
   of the substructure with the intention of depicting the overall
   shape. The right panel is a projection of the substructure onto
     the Galactic plane. The arrows show the total velocity in the
   Galactic rest-frame. The Sun is marked as a star at the centre,
   whilst the Sun's motion is marked by an arrow in magenta. A sphere
   of radius 2 kpc (which is a crude representation of the Galactic
   bulge), as well as a grey sheet representing the Galactic plane,
   are shown to give a sense of the scale and position of the
   substructure in relation to the familiar Galactic
   landscape. A triad of velocity vectors of scale 300 kms$^{-1}$
   is shown in the bottom left corner.} 
 \label{fig:hotstuff}
 \end{figure*}
 \begin{figure*}
 \begin{center}
 \includegraphics[width=0.9\textwidth]{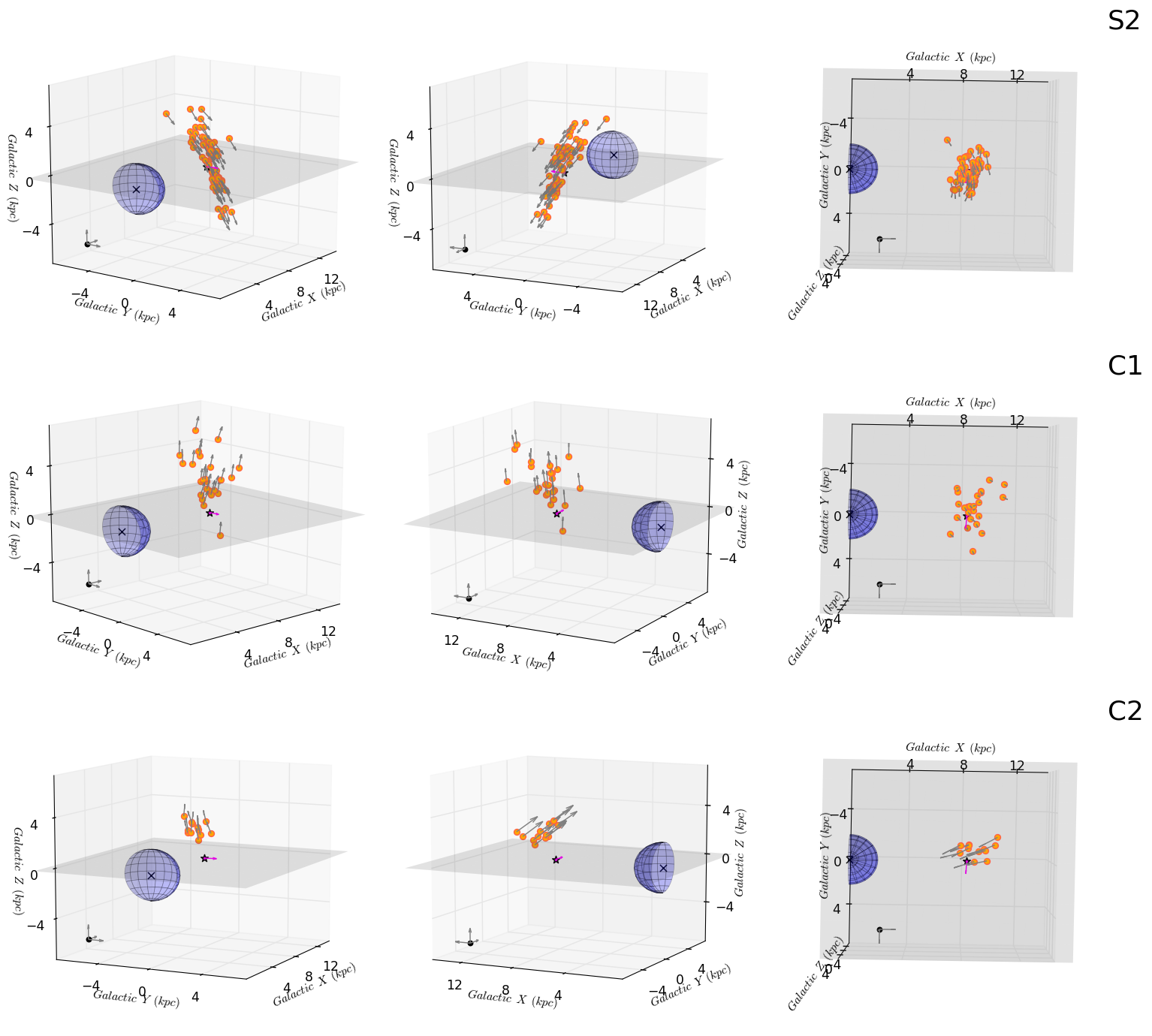}
 \end{center}
 \caption{As Fig.~\ref{fig:hotstuff}, but for the stream S2 and the
   two moving clumps (C1 and C2).}
 \label{fig:coldstuff}
 \end{figure*}
\begin{figure*}
 \begin{center}
 \includegraphics[width=0.9\textwidth]{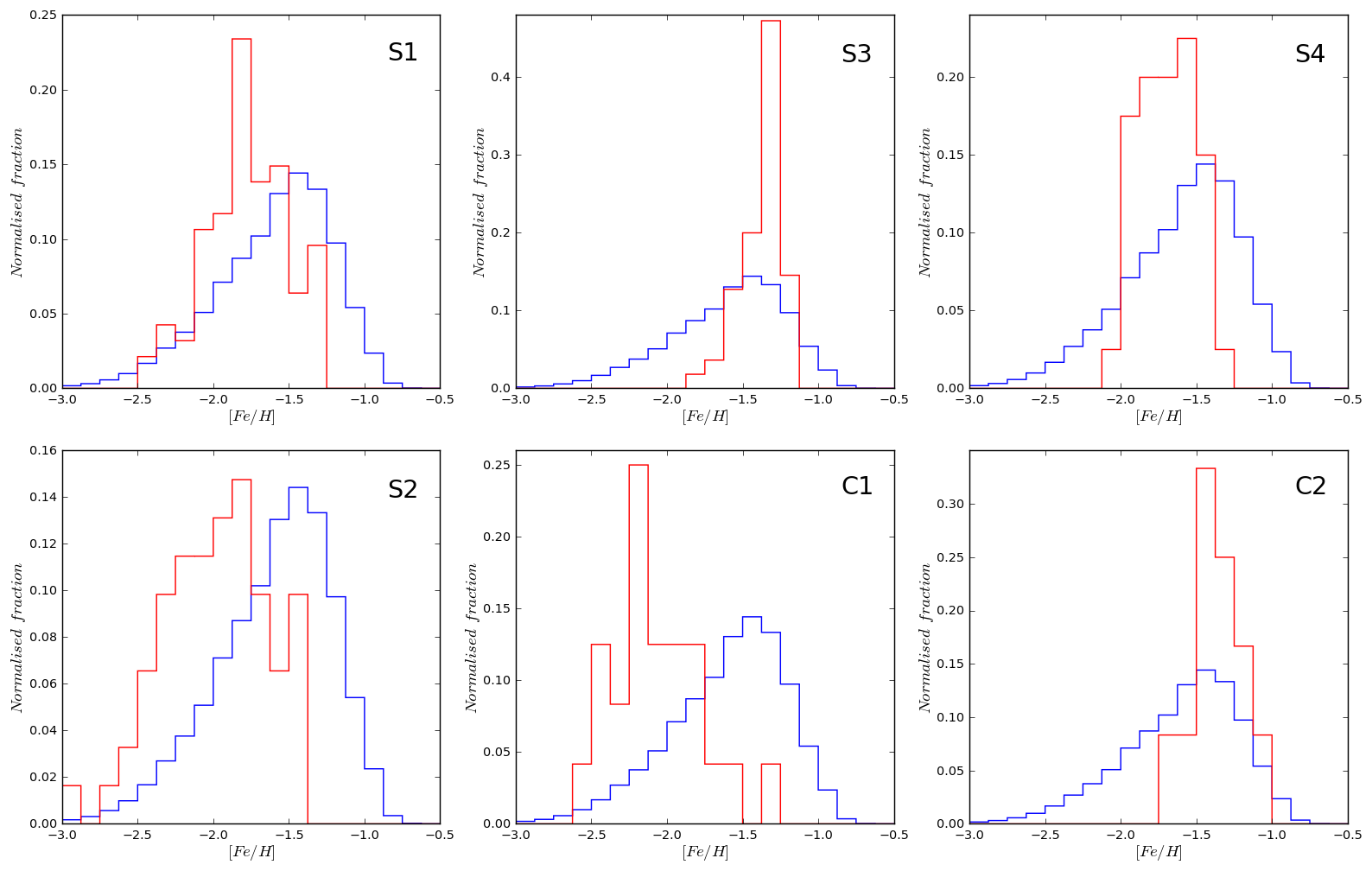}
 \end{center}
\caption{The metallicity distribution function for the six
  substructures is shown in red, whilst the blue is the entire halo
  sample for comparison. Note that the substructures are narrower in
  metallicity than the entire halo, which is consistent with
  expectations.}
\label{fig:metaldfs}
\end{figure*}

 \section{Method}

 \subsection{Sample}

 Our starting sample is the crossmatch between Gaia data release 1
 (DR1), the Sloan Digital Sky Survey data release 9 (PhotoObjAll for
 the photometric and sppParams for the spectroscopic), APOGEE, LAMOST
 DR2 and RAVE-on~\citep[see e.g.,][]{An14,Lu15,Ca17,Ku17}. There are
 466\,414 stars in this sample with five-dimensional phase space
 information.  The sample contains MSTO stars and BHB stars, which can
 be extracted using methods similar to Sections 3.1 and 3.2 of
 \citet{Wi17}.  The MSTO stars are extracted using the cuts:
 extinction $\epsilon_r < 0.5$, $g,r,i$ magnitudes satisfying $14 < g
 < 20$, $14 < r < 20$, $14 < i < 20$, $0.2 < (g-r)_0 < 0.8$ with
 surface gravity $3.5 < \log g < 5.0$ and effective temperature $4500
 < T_{\rm eff} < 8000$. The BHB stars are chosen from $-0.25 < (g-r)_0
 < 0.0$, $0.9 < (u-g)_0 < 1.4$ with spectroscopic parameters
 satisfying $3.0 < \log g < 3.5$ and $8300 < T_{\rm eff} < 9300$. We
 apply a set of quality cuts to both the photometric and spectroscopic
 data to remove stars with uncertain measurements as well as stars
 with a heliocentric radial velocity error $> 15$ kms$^{-1}$ and a
 heliocentric distance $> 10$ kpc. The cuts cause the sample to be
 reduced to 245\,316 in size with 245\,078 MSTO stars and 238 BHB
 stars. The median heliocentric radial velocity error is $2.9$
 kms$^{-1}$ and the median proper motion error is $17.8$
 kms$^{-1}$. Parallaxes can be obtained via the formulae in
 \citet{Iv08} for MSTOs (using spectroscopic metallicities) and in
 \citet{De11} for BHBs to give full six-dimensional phase space
 information. For the MSTOs that comprise the bulk of the sample, mean
 distance error scales linearly with distance and reaches $\sim 1$ kpc
 at a distance of 4.5 kpc. The mode of the distance error for the
 whole MSTO sample is $\sim 0.47$ kpc.

 Velocities in the Galactic rest-frame are resolved in the cylindrical
 polar coordinate system to give ($v_R,v_\phi, v_z$).  From the
 histogram in the ($v_\phi, [{\rm Fe/H}])$ plane in
   Fig.~\ref{fig:figone}, we see a reasonably clear separation of
 the halo population from the thin and thick disk populations. We
 define a polygon (converted from a contour) representing each
 population, and then calculate the distance of each star from two
 contours (one representing the halo, the other representing the thin
 and thick disks). This enables us to classify each star as either
 halo or disk.  For the halo stars, we perform a Gaussian fitting
 decomposition based on the metallicity, and then subdivide the halo
 group into the relatively metal-rich, and the relatively metal-poor
 halo. As the result of the Gaussian decomposition, the division
 occurs at [Fe/H] $\approx -1.65$.  Our sample then comprises 181\,574
 disk stars (green), 40\,293 relatively metal-rich halo stars (blue),
 and 23\,449 relatively metal-poor halo stars (red), as shown in
 Fig.~\ref{fig:figone}.

 This subdivision of the stars into disk and halo groups is crude, but
 we only wish to use it to demonstrate that the sequence from thin and
 thick disk through metal-rich halo to metal-poor halo is one of
 increasing substructure. This is evident from Fig.~\ref{fig:lumpy} in
 which the logarithmic contours of the velocity distribution in the
 ($v_R, v_\phi$) plane moves from smoothness to raggedness with
 increasing numbers of outliers and subgroups.  Some of this effect is
 statistical in origin as there are between 4 and 8 times fewer stars
 in the halo populations. However, some prominent pieces of halo
 substructure can be picked out by eye, and so some of the effect is
 real. Accordingly, we proceed to develop a systematic way of
 identifying the substructure.

 \subsection{Detection}

 Henceforth, we use the entire halo sample (the blue and red
 distributions in Fig.~\ref{fig:figone}). We first develop a smooth
 underlying background model, which is then used as the global density
 estimator against which substructure is identified. Using
 Galactocentric velocities resolved with respect to cylindrical polar
 coordinates ($v_R, v_\phi, v_z$), we fit a basic Gaussian Mixture
 model from the {\tt Scikit-learn} \citep{Pe11} python software
 package~\footnote{http://scikit-learn.org}. Note that if we use too
 many Gaussian components, some of the actual signals from genuine
 substructures get diluted by some of the fitted Gaussians.  To avoid
 such dilution, we decide to use considerably less Gaussian components
 than the estimate of the number of components obtained from 
 minimization of the Akaike Information Criterion or AIC test 
 (56 components).  We ensure that each of the fitted Gaussians has a 
 width wider than $150$ kms$^{-1}$ on each axis to avoid including small 
 scale features in our velocity distribution model.  We find that 10 
 Gaussians provide a reasonable description of the velocity space for 
 the halo stars, as shown in Fig.~\ref{fig:residuals}.  The data, 
 together with the superposed Gaussians are shown in the left panels, 
 whilst the smooth model and residuals are shown in the middle and 
 right. It is evident that there is substructure, and much of it 
 corresponds to prominent clumps in Fig~\ref{fig:lumpy}.

 Next, we look for significant overdensities over the Gaussian Mixture
 model.  We measure the local density of each star in our data, and
 compare this to the density value predicted by the smooth model.  We
 do this by carrying out a $k$-nearest neighbours search with $k = 5$
 (or 6 including the star itself). Using {\tt Scikit-learn}, we obtain
 the radius $r_5$ required to encounter the $k=5$ nearest neighbours
 and hence an estimate of the local density.  The probability of the
 star's location in the 3-dimensional velocity space is predicted by
 the Gaussian Mixture model.  We multiply this by the sample size and
 the volume of the sphere with radius $r_5$ to give the expected
 number.  We assume Poisson sampling and from the expected number of
 stars in the sphere, we compute the tail probability of having 6
 stars (5 neighbours and the star itself) in the sphere given this
 distribution. We then convert the tail probability to the number of
 sigma.

 We use any stars with significance $> 4$ as the \lq\lq seeds" for
 finding an overdensity in our 3-dimensional velocity space. First, we
 classify these seeds by the Friends-of-Friends method -- that is, any
 seeds that are close to each other ($<30$ kms$^{-1}$ radius sphere)
 are considered as the same group.  For each seed, we then take all
 stars within a spherical volume of radius $35$ kms$^{-1}$ around the
 seed. During this process, we discard any seeds and corresponding
 stars if there exist less than 5 stars within this spherical volume.
 We classify the stars around the seeds by using the Nearest
 Neighbours Classification from {\tt Scikit-learn}. This stage is
 necessary because there are cases in which a star is picked up by
 more than one seed.  So we train the classifier using the classified
 seeds, and then perform a distance-weighted $k$ neighbours
 classification ($k = 3$) for the stars around the seeds.  The weight
 here is the inverse of the distance. This gives us a list of
 candidates.

 Now, we find the centre of each group in our 3-dimensional space. The
 measured number of group members is the number of stars in the
 ellipsoid in velocity space occupied by the group.  This ellipsoid
 has a volume $4\pi abc/3$, where ($a,b,c$) is the extent of the group
 in each axis.  The expected number of field stars in the volume
 ellipsoid is then the probability predicted by the Gaussian Mixture
 Model at the central location multiplied by the data size and by the
 volume. The Poisson uncertainty is the square root of the expected
 number. This provides us with a crude measure of the significance of
 each substructure.

 We will provide the list of substructures elsewhere, but here we
 describe the six most significant pieces of halo substructure, which
 is $\sim20$ per cent of the detected potential candidates with
 $\sigma > 4$. They are labelled S for stream or shell-like
 substructures and C for clusters or moving groups. The locations of
 the stars in velocity space belonging to the substructures are shown
 in Fig.~\ref{fig:figvd}. Note that, as the stars lie within the SDSS
 footprint, proper motions contribute mainly to the radial $v_R$ and
 azimuthal $v_z$ components, whilst the line of sight velocities
 contribute mainly to $v_z$. This causes kinematic features to
 appear colder in $v_z$ than in the other two directions which
 are more affected by errors.  The two largest substructures in terms
 of the number of member stars are S1, coloured blue, with 94
 identified members ($\sigma = 8.94$) and S2, coloured red, with 61
 members ($\sigma = 8.95$). Just behind them in terms of the number of
 member stars are: S3, coloured magenta, with 55 members ($\sigma =
 8.41$) and S4, coloured green, with 40 ($\sigma = 8.49$)
 members. There are also two clumps or moving groups: C1, coloured
 brown, with 24 members ($\sigma = 8.46$) and C2, coloured pale blue,
 with 12 members ($\sigma = 18.66$).  Table~\ref{table:params}
 provides the median, mean absolute deviation and dispersion for
 kinematical and spectroscopic quantities of the substructures. A list
 of stars in the substructures is available electronically from the
 authors.

 \begin{table*}
 \begin{center}
   \begin{tabular}{lrrrrrrrrc}
 \hline \hline 
 \multicolumn{1}{c}{Name} & \multicolumn{1}{c}\null & 
 \multicolumn{1}{c}{[Fe/H]} &
 \multicolumn{1}{c}{$\log g$} & \multicolumn{1}{c}{$T_{\rm eff}$} &
 \multicolumn{1}{c}{$(X,Y,Z)$} & \multicolumn{1}{c}{$(v_R,v_\phi,v_z)$}
 & \multicolumn{1}{c}{$(U,V,W)$} &\multicolumn{1}{c}{KE} & \multicolumn{1}{c}{L}
 \\ \multicolumn{1}{c}\null &\multicolumn{1}{c}\null &
 \multicolumn{1}{c}{\null} & \multicolumn{1}{c}{\null} &
 \multicolumn{1}{c}{(K)} & \multicolumn{1}{c}{(kpc)} &
 \multicolumn{1}{c}{(kms$^{-1}$)} & \multicolumn{1}{c}{(kms$^{-1}$)} & (km$^2$s$^{-2}$) & (kpc kms$^{-1}$)\\
\hline \null & Median& -1.78 & 3.96 & 6073.2 & (8.1,-0.2,2.8)& (44.8,-313.8,-42.7)& (32.9,-322.6,-42.7) & 55872 & 2911 \\ 
 S1 & MAD & 0.19 & 0.21 & 265.7 & (0.4,0.9,1.1) & (38.4,75.7,21.3) & (25.8,65.4,21.3) & 22621 & 519 \\ 
 \null & Dispersion & 0.27 & 0.25 & 335.2 & (0.8,1.2,1.9) & (56.2,93.4,26.5) & (39.2,86.5,26.5) & 26815 & 822\\ \hline
 \null & Median& -1.91 & 4.00 & 5847.0 & (8.7,0.3,0.8) & (8.9,160.2,-249.9) & (-1.2,164.9,-249.9) & 46048 & 2632 \\
 S2 & MAD &0.26 &0.26& 392.4 & (0.3,0.6,1.5)&(19.7,12.1,11.7)& (17.1,10.2,11.7) & 2527 & 127 \\
 \null & Dispersion & 0.35 & 0.32 & 473.8 & (0.6,1.0,2.0) & (28.2,16.8,17.5) & (35.6,15.5,17.5) & 4962 & 233 \\ 
\hline
\null & Median &
-1.34 & 4.05& 6114.3 & (8.6,0.5,3.5)&(50.6,-245.5,206.7) & (77.3,-257.0,206.7)&59857 & 2901\\
S3 &MAD &
0.08 & 0.20 & 196.8 & (0.6,1.0,1.0)& (67.3,30.5,19.0)&(60.1,40.6,19.0) & 11230 & 448\\
\null & Dispersion & 0.13 & 0.30 & 334.2 & (1.1,1.5,1.9) &(80.1,58.7,22.4)&(66.1,61.5,22.4)&17604 & 792\\ \hline
\null & Median &
-1.70 &3.83 & 6144.1& (8.5,0.6,4.1)&(4.0,-250.5,157.7)&(14.7,-262.7,157.7)&49617&2755\\
S4&MAD&
0.14 &0.14 &203.8 & (0.5,1.2,1.8)&(68.2,43.8,18.5)&(41.3,38.3,18.5)&11436&351\\
\null&Dispersion&
0.19 & 0.28 & 290.7& (1.0,,1.7,3.6)&(84.4,54.3,22.2)&(61.6,56.2,22.2)& 14708 &719\\\hline
\null&Median&
-2.11 & 3.85 & 6081.4 & (8.7,-0.8,2.5)&(32.0,32.3,271.6)& (33.7,29.4,271.6)&38235&
2311\\
C1&MAD&0.16 &0.25 &285& (0.4,1.0,1.1)&(8.5,18.6,11.2)&(10.7,20.4,11.2)&2502&107\\
\null&Dispersion&
0.29&0.31&360& (0.9,1.4,1.6)&(17.8,26.9,15.8)&(18.2,28.3,15.8)&4807&191\\\hline
\null&Median&
-1.39& 4.05&5998.7& (8.9,-1.0,2.1)&(-337.5,75.4,295.0)&(-322.7,121.0,295.0)& 105895&3498\\
C2&MAD&
0.12&0.21&258.9&(0.6,0.3,0.3)&(13.1,16.9,8.2)&(15.3,27.8,8.2)&3319& 147\\
\null&Dispersion&
0.17&0.25&320.1&(0.85,0.5,0.6)&(16.2,23.9,10.0)&(18.6,32.3,10.0)&6631&293\\
\hline
   \end{tabular}
 \end{center}
  \caption{The median, median absolute deviation and dispersions in
    spectroscopic and kinematic properties of the six substructures.}
  \label{table:params}
\end{table*}

\begin{figure}
\begin{center}
\includegraphics[width=0.5\textwidth]{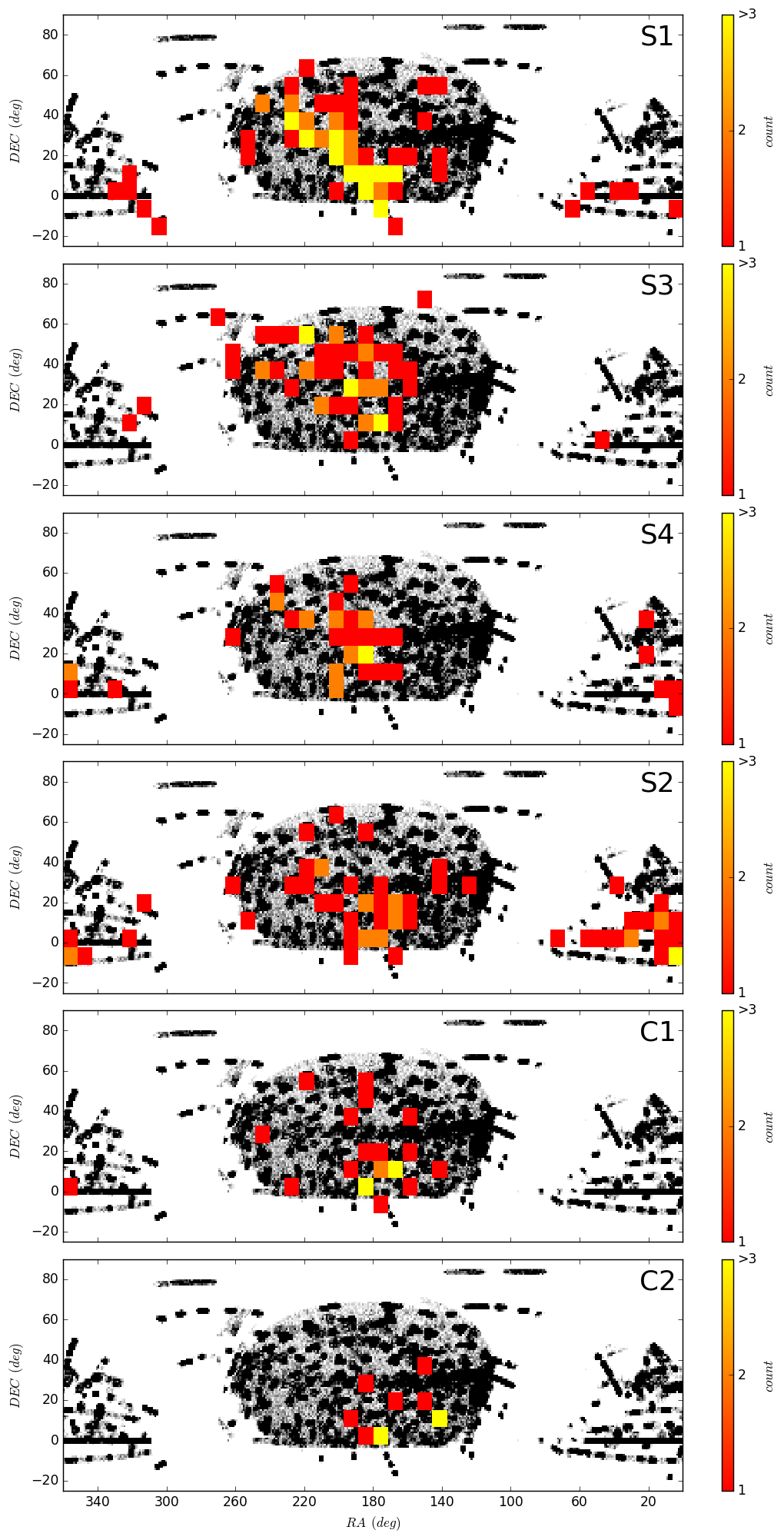}
\end{center}
\caption{The location of the stars belong to the substructures are
  shown in the plane of right ascension versus declination. The pixel
  size is 8.5\,deg on each side.}
\label{fig:equatorial}
\end{figure}

\section{Candidates}

\subsection{The Hotter Substructures: S1, S3 and S4}

Fig.~\ref{fig:hotstuff} shows the discovery panels for the three
hotter substructures. For each, we provide two views of the morphology
in the left and middle panels, as well as a projection onto the
Galactic plane on the right. The metallicity distribution function of
each substructure is compared against that of the full halo sample in
Fig.~\ref{fig:metaldfs}.

S1 is a large piece of halo substructure, containing 94 member
stars. The members correspond to an obvious narrow tail-like
overdensity in the ($v_\phi, v_z$) velocity distribution in
Figs~\ref{fig:lumpy} or \ref{fig:figvd}, visible by eye. The medians
of the positions of the stars provide a location of ($X,Y,Z$) = ($8.1,
-0.2, 2.8$) kpc, so the structure lies just beyond the Sun's
location. It has a substantial extension in both $Y$ and $Z$ as
indicated by the median absolute deviations of $\sim 1$ kpc, so it is
distended vertically and azimuthally. Therefore, the spatial
configuration is shell-like, pirouetting around the Sun's location.
The vertical or $v_z$ velocities are tightly constrained around a
median of $-42.7$ kms$^{-1}$ with a median absolute deviation of
$21.3$ kms$^{-1}$.  The structure is counter-rotating with a median
$v_\phi$ of $-313.8$ kms$^{-1}$. The median radial velocity $v_R$ is
$44.8$ kms$^{-1}$ with a comparatively large median absolute deviation
of $38.4$ kms$^{-1}$, mainly caused by the extent of the structure.
It is natural to inquire whether this is a diminutive analogue of the
shell-like features seen in elliptical galaxies~\citep{He87} or in the
Milky Way halo~\citep{Ro04,Be07}. However, shells are known to be
associated with radial infall of galaxies or
clusters~\citep[e.g.,][]{Am15,He15,Po17}, whereas the strongly
counter-rotating nature of the substructure indicates that the
progenitor orbit has high angular momentum. We will elaborate on the
true nature of this structure in the next Section.

The detection algorithm used to identify substructures is based on
kinematics alone. However, in all our presented substructures, it is
possible to identify clumpiness in configuration space and in
chemical properties.  Fig.~\ref{fig:metaldfs} shows the metallicities
of the S1 stars in red are much spikier than the halo metallicity
distribution function in green. They have a median metallicity of
$-1.78$ with a narrow median absolute deviation of $0.19$, making this
a convincing detection.

S3 and S4 share some similarities with S1 in that the radial and
azimuthal velocity distributions are broad, but the vertical velocity
distribution is narrower, suggesting a highly inclined orbital
plane. S3 and S4 are more obviously stream-like, as the stars are
moving along the extent of the structure, whereas S1 moves almost
perpendicularly.  All three substructures are on retrograde
orbits. They all lie just beyond the Solar position, though the
preponderance of substructure here is a selection effect of the
SDSS-Gaia catalogue. The stars belonging to both S3 and S4 are tightly
clustered in metallicity with median values of $-1.34$ and $-1.70$
respectively. Although S3 and S4 occupy similar region in the
3-dimensional velocity space, they show clear deviation in their
metallicity distribution as well as in their $v_z$ distribution which
suggest they are separate substructures. This has been further checked
by the Gaussian fitting decomposition on the 4-dimensional space
(3-dimensional velocity components and the metallicity) which shows
the separation between two substructures more clearly. Notice that S3
is comparatively metal-rich and is visible by eye as a distortion in
the outermost contours of the velocity distribution of the metal-rich
halo in the middle panel of Fig.~\ref{fig:lumpy} at $(v_\phi,v_z)
\simeq (-250,200)$ kms$^{-1}$. There is also a possibility that S3 and
S4 are not fully distinct substructures. Despite their different
metallicity distributions, their close overlap in velocity space
(Fig.~\ref{fig:figvd}) and similar spatial distribution (middle and
bottom rows of Fig.~\ref{fig:hotstuff}) suggest a possibility of a
single large substructure with some internal metallicity variations
being torn apart over time.

\subsection{The Colder Substructures: S2, C1 and C2}

The top row of Fig.~\ref{fig:coldstuff} shows the discovery panels for
substructure S2 comprising 73 member stars, which has the
characteristics of a halo stream.  S2 corresponds to an obvious
overdensity in the ($v_\phi,v_z$) velocity distribution. It can be
seen as an underhanging blob of stars in the lower rightmost panel of
Fig.~\ref{fig:lumpy} at $(v_\phi,v_z) \simeq (160,-250)$
kms$^{-1}$. The member stars also comprise a tight grouping in the
($v_R, v_\phi$) and ($v_R, v_z$) planes. The coldness of this
substructure in velocity space is emphasised by the narrow velocity
distributions. The median absolute deviation in ($v_R, v_\phi, v_z$)
are ($19.7,12.1,11.7$) kms$^{-1}$, though these are of course averages
over the spatial extent of the stream and so are not indicative of the
velocity dispersion or the size of the progenitor.

The median values of the spatial coordinates are ($X,Y,Z$) = ($8.7,
0.3, 0.8$) kpc, so that this substructure is again close to the
Sun. Nearby objects have the highest proper motions and stand out from
the bulk of the stars in the catalogue, so it is not surprising that
our detection is more sensitive to the substructures close to the
Solar radius. As Fig.~\ref{fig:coldstuff} shows, S2 is a stream
plunging through the Galactic disk, moving on a nearly polar orbit.
The fact that the stream is aligned along the velocity vectors of the
stars, as is natural for a stream, adds confidence to our
detection. The stars have a median metallicity [Fe/H] of $-1.91$ and a
median absolute deviation of $0.26$. As is clear from
Fig.~\ref{fig:metaldfs}, the metallicity distribution function of the
substructure is poorer and narrower than the stellar halo as a whole.

In fact, S2 lies at a very similar location in velocity space as 4
stars belonging to the halo stream identified in Hipparcos data by
\citet[][see especially the upper panels of their Fig. 2]{He99}. Their
stars are clumped in ``integrals of motion space'', while the two
structures have no direct member stars in common, presumably due to
the use of different dataset. The relationship of S2 with the stream
of \citet{He99} will be discussed in detail elsewhere. As the
associated substructure has been identified both in velocity space and
in ``integrals of motion space'', it provides an interesting test case
for assessing the advantages and disadvantages of each search arena
and algorithm.
  
The middle and bottom rows of Fig.~\ref{fig:coldstuff} show panels for
the two clumps C1 and C2. These comprise 24 and 12 members
respectively, and so are less substantial and extensive than
S1-S4. Their velocity histograms are very narrow with the vertical
velocity distribution being the coldest. The structures are tightly
confined in space and in metallicity. The median metallicity [Fe/H] of
C1 is $-2.11$, making it the most metal-poor of all our substructures,
whilst C2 has a median metallicity of $-1.39$ (see
Fig.~\ref{fig:metaldfs}).

\subsection{Distribution on the Sky}

The locations of the stars in right ascension and the declination for
all the substructures are shown in Fig.~\ref{fig:equatorial}. Notice
that the substructure are difficult to discern, with the exceptions of
S1 and S4. In general, the substructures are both nearby and extended,
so their member stars are scattered across the sky. The stream S2 is
hard to make out, as it is traversing the Galactic
disk. Fig.~\ref{fig:equatorial} vindicates the power of kinematic
searches, as the substructures would be nearly impossible to identify
any other way.

\begin{figure*}
\begin{center}
\includegraphics[width=0.85\textwidth,angle=0]{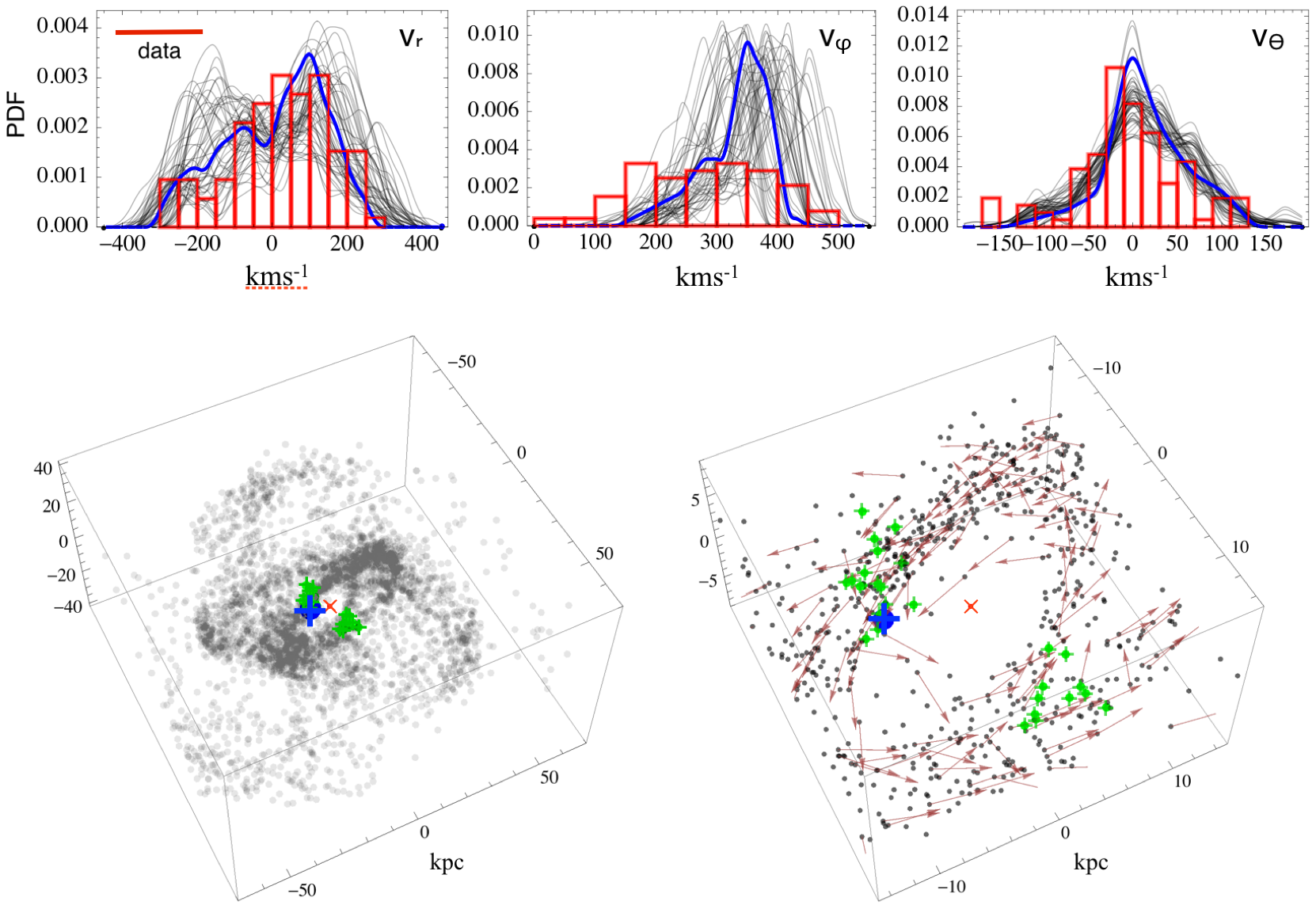}
\includegraphics[width=0.85\textwidth,angle=0]{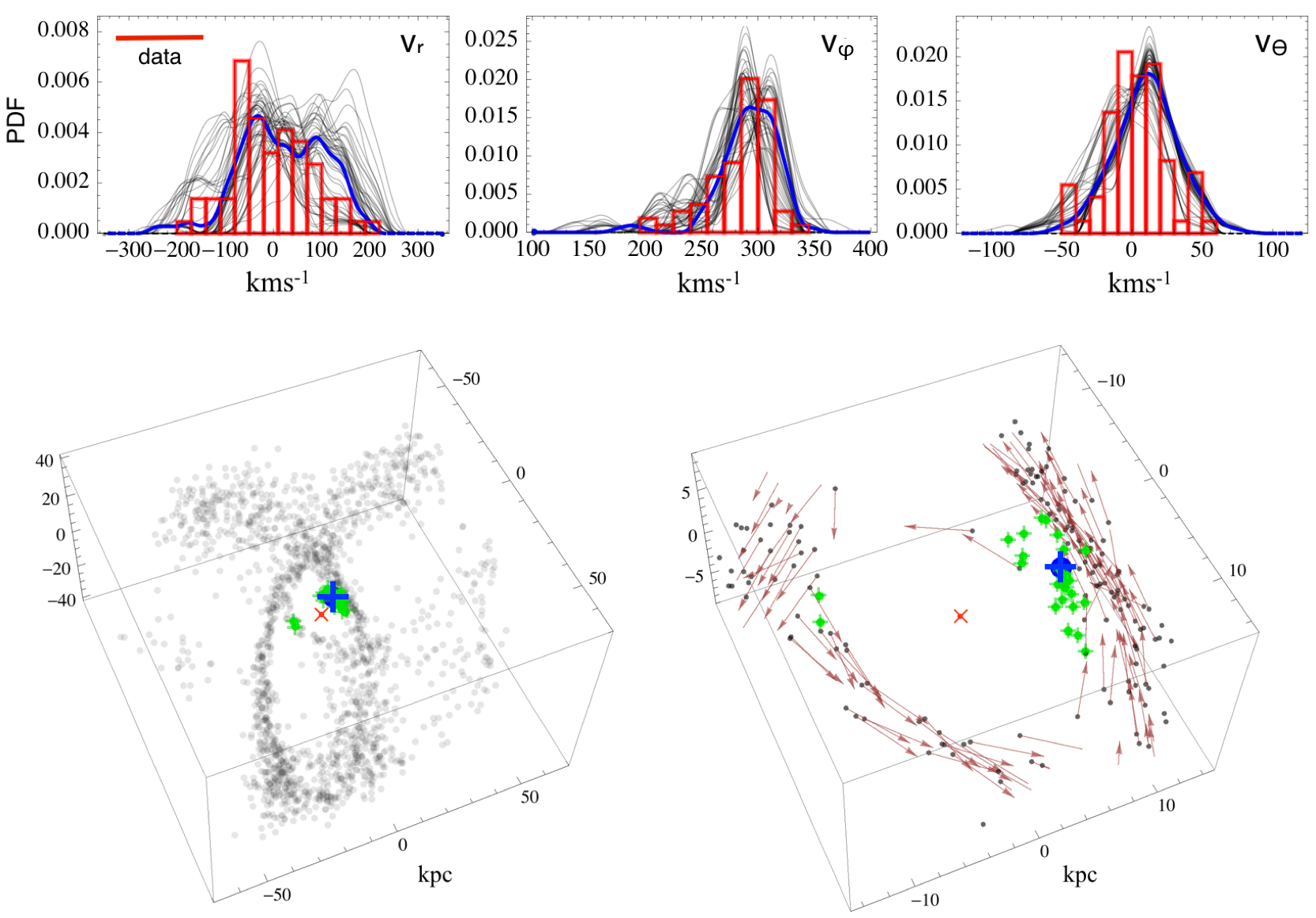}
\end{center}
\caption{Matches to substructures S1 (upper two rows of panels) and S2
  (lower two rows of panels) in the library of Amorisco (2017). For
  each substructure, the observed kinematics (red histograms) is
  compared with the chosen model in the upper trios of panels. There,
  thin grey lines illustrate the debris' kinematics corresponding to
  different viable positions of the Sun. The thick blue line
  identifies the best-fitting model, corresponding to the best Sun's
  position. The lower-left panels illustrate the three-dimensional
  structure of the simulated tidal debris. Grey points are model
  particles and the red X symbol identifies the Galactic centre. Green
  $+$ symbols identify the Sun's positions corresponding to the
  kinematic distributions shown in the upper trio of panels. The best
  Sun's position is displayed with a large blue $+$ symbol. The panel
  in the lower-right is a zoomed version that best shows the position
  of the simulated debris material with respect to the selected Sun's
  locations. Symbols are as in the lower-left panels. Additionally, a
  fraction of the model particles are accompanied by their velocity
  vectors, to illustrate the debris kinematics.}
\label{fig:fits}
\end{figure*}

\section{Interpretation}

We use the library of accretion events created by \citet{Am17} to find
model analogues for the two largest substructures S1 and S2. The
library uses minor merger N-body simulations to study how stellar
material is deposited onto the host. Both host and infalling
satellites are assumed to have spherical Navarro-Frenk-White
profiles~\citep{Na97}, meaning that the Milky Way disk is not properly
accounted for. The disk is not expected to cause substantial
additional satellite disruption in satellites with total masses
$M_{sat}\gtrsim10^9 M_\odot$ \citep{ED10,GK17}, but can alter the
debris' orbits and it does increase the speed of the phase-mixing
process \citep[e.g.,][]{HW99}. This initial exploration neglects these
effects. Specifically, we search for accretion events that result in
substructures located close to the Solar radius and that provide
reasonable matches to the velocity histograms. To do so, we use
spherical polar coordinate, ($v_r, v_\theta, v_\varphi$), defined by the
mean angular momentum vector of the substructure itself. Therefore,
$v_\varphi$ refers to rotation in the mean orbital plane, while the
scatter in $v_\theta$ is a proxy of the structure's hotness.  By
automating this, we can explore a large number of models in the
library with a variety of mass ratios ($-2\lesssim \log
M_{sat}/M_{halo}\lesssim-0.5$), infall circularities ($0.2<j<0.8$,
where $j$ is the ratio between angular momentum and the maximum
angular momentum at the same energy) and infall times. For each model
we look for matches by considering thousands of possible Sun's
locations, together with slightly different mass and length
normalisations. The former exploits the lack of a stellar disk in the
models, the second explores the possible scatter in the values of the
Galaxy's mass and concentration at the time of infall.

Despite the size of this library, there do not seem to be that many
models that fit reasonably when actually compared with the histograms
of S1. For a randomly picked Sun's location, most model structures
feature the presence of multiple phase-space wraps, resulting in
sharply double peaked $v_r$ distributions, with average close to zero
or $\langle v_r\rangle \approx 0$.  Instead, S1 is characterised by a
broad and unimodal $v_r$ distribution, which contains $v_r=0$ and is
not double peaked. To reproduce this, the Sun is required to lie very
close to the pericenter of the debris' orbit. Among those models for
which feasible locations for the Sun can be found, we illustrate one
of the most successful in the upper panel of Fig.~\ref{fig:fits}, in
which green points display a selection of feasible Sun's
locations. Each of these produce the velocity distributions plotted as
thin black lines in the upper panels. The best choice for the Sun's
location is shown in blue.  The corresponding blue velocity
distributions reproduce most of the features of the velocity
histograms, though the match to the $v_\varphi$ distribution is
poor. It corresponds to a virial mass ratio of $M_{sat}/M_{halo}=1:20$
at infall, implying that the progenitor had a starting mass of
$\approx 2 \times 10^{10} \msun$ at infall time $\approx 10$ Gyr ago,
for a circularity at infall of $j=0.8$.

As shown in the lower plot of the upper panel of Fig.~\ref{fig:fits},
S1 is identified as a stream in an advanced state of
disintegration. The Sun appears to lie within the stream's wraps,
while these are at pericenter. The quite advanced state of phase
mixing and partial superposition of multiple stream wraps helps in
reproducing the broad $v_\varphi$ distribution, although the model
distributions still appear to remain somewhat tighter than suggested
by the data. The stream does indeed pirouette around the Sun, but the
substructure S1 is not a shell. In fact, its angular momentum is still
high, as permitted by the low initial virial mass ratio. The $v_r$
distribution encompasses $v_r = 0$, but it does so while the Sun is
close to the stream's pericenter rather than to the apocenter as in a
more classical shell. The fact that one of the velocity distributions
is poorly fit does mean that our conclusions regarding the properties
of the progenitor of S1 are preliminary. It may be that we have a
restricted view of S1 owing to the incompleteness of our sample,
though integration of the orbits of the stars does not reveal a
connection to other known substructures.
  
The $v_r$ histogram of the substructure S2 has a similar
distinguishing property, implying that the Sun's preferred position is
again very close to pericentre. The main difference is that the
dispersions are smaller, which drives the model to lower mass ratios,
and therefore to less phase-mixed morphologies. The lower panel of
Fig.~\ref{fig:fits} illustrates a model that provides a good match: it
has a virial mass ratio of $1:100$ at infall, implying that the
progenitor had a starting mass of $\approx 5 \times 10^{9} \msun$ at
infall time $\approx 11$ Gyr ago, and an initial circularity of
$j=0.5$. However, a number of models that are close in parameter space
can also roughly reproduce the features of the substructure. For
example, one can trade a slightly higher initial mass ratio ($\approx
1:50$) for a somewhat later infall time ($\approx 8$ Gyr) or a
marginally higher angular momentum.  These coupled changes can
compensate each other, without affecting much the degree of the
stream's phase mixing, and therefore its kinematic properties.

As shown by the analysis above, a significant variety of models have a
pericentric distance that is comparable with the Sun's radius. These
are models with comparatively old progenitors, which helps them to
fall deeper in the Milky Way halo, but not overly massive, which would
instead cause excessive dynamical friction and phase mixing.  Despite
the limits of the models we adopted, it is clear that the progenitors
of both S1 and S2 belong to this class.

The inferred total masses of S1 ($\approx 10^{10} \msun$) and S2
($\approx 5 \times 10^9 \msun$ are about a factor of 10 smaller than
the total mass of the Large Magellanic Cloud. According to the
abundance matching of \citet{GK14}, these correspond to stellar masses
between $10^6$ and $10^7 \msun$, and so are comparable to present-day
objects like the Fornax dwarf spheroidal. The stellar masses inferred
for S1 and S2 through the mass-metallicity relation of \cite{Ki13} are
$10^{5.7} \msun$ and $10^{5.3} \msun$, which are somewhat lower by
factors of $5$ to $15$. However, this does not take into account the
redshift evolution of the mass-metallicity relation~\citep[see
  e.g.,][]{Ma16}, which though uncertain may remove these
inconsistencies entirely.  In addition, there is substantial scatter
in both abundance matching, the mass-metallicity relation and the data
of \citet{Ki13}.  Hence, metallicity and kinematics appear to be
painting a broadly consistent picture.

Nonetheless, there are some clear shortcomings to our
methodology. First, we did not use the footprint of the SDSS-Gaia
survey and so this weakens our claim to a proper comparison with the
data. Secondly, the proper motion errors are not known on a star by
star basis, though on average they are reckoned to be $\sim 2$ mas
yr$^{-1}$. The effect of the proper motion errors is to broaden the
distributions in the angular coordinates especially and this may
partially explain our failure to reproduce the broadness of the
$v_\varphi$ distribution for S1. Finally, the underlying galaxy models
used to generate our substructure library are spherical and so
somewhat idealised.

\section{Conclusions}

We have devised a method to search algorithmically for
substructure. We model the distribution of the underlying smooth
component as a Gaussian Mixture model.  We use this to identify
enhancements against the background, by comparing the local density
around any star with the prediction from the Gaussian Mixture model
and thence computing the significance. Stars with significance greater
than 4 are then grouped by a Friends-of-Friends algorithm to give
substructures.  In our application, the underlying smooth component is
the velocity distribution of the stellar halo, and we were seeking
kinematically coherent substructures that are the residue of
long-disrupted dwarf galaxies.

Our method has a number of advantages. First, the entire algorithm is
very fast. For the halo samples studied here (63\,742 stars),
substructures can be identified and their significance computed in
$\sim 100$s.  It is estimated that there will be $2 \times 10^7$ halo
stars in Gaia Data Release 2~\citep{Ro12}, so the algorithm remains
competitive and feasible in the face of the much larger datasets
expected shortly. Secondly, the algorithm is easily adapted to
different search spaces. Here, we chose to search only in velocity
space and use any metallicity data as confirmation. However, it would
have been easy to add extra dimensions in chemistry (such as
metallicity or abundances) and search in a chemo-dynamical space.
Alternatively, we could have applied the algorithm in action or
'integral of motion' space.

We implemented the new algorithm on a sample of stars extracted from
the SDSS-Gaia catalogue~\citep[see e.g.][]{De17}. This uses Sloan
Digital Sky Survey (SDSS) photometry as the first epoch for sources in
Gaia DR1. When cross-matched with available spectroscopic surveys,
such as RAVE, APOGEE or LAMOST, we obtain the line-of-sight velocities
and metallicity.  By photometrically selecting main sequence turn-off
stars or BHB stars, for which distance estimators are available, we
construct a sample of 245\,316 stars with full phase space
coordinates. The velocity distributions show a strong trend of
increasing substructure with diminishing metallicity. The most
metal-poor stars ([Fe/H] $< -1.65$) exhibit abundant substructure in
their velocity distributions. Some of the substructures are visible by
eye.

Our new algorithm enabled us to identify six new substructures in the
local stellar halo.  The most substantial (S1) is a stream in an
advanced state of disruption just beyond the Solar radius. The Sun is
located close to the pericentre of multiple wraps, giving rise to a
broad distribution in two of the velocity components. This is the
relic of an old accretion event in which a satellite was engulfed on a
retrograde orbit. Modelling suggests that the progenitor was
relatively massive at $\approx 2 \times 10^{10} \msun$ at infall time
$\approx 10$ Gyr ago.  The next most substantial (S2) is a stream,
though it is more intact.  Again, it is located close to the Solar
radius, but is plunging through the Galactic disk.  It has
characteristic stream kinematics, with the velocity vectors of the
stars aligned with the elongation of the substructure. The cold
velocity distributions suggest that the progenitor was less massive --
at most perhaps $\approx 5 \times 10^{9} \msun$ at infall time
$\approx 11$ Gyr ago.  The stars belonging to these substructures are
clustered not just kinematically but also chemically, which adds
confidence to the detections.  Abundance matching suggests that both
S1 and S2 correspond to galaxies with stellar masses between $10^6$
and $10^7 \msun$. This is comparable to the largest dwarf spheroidal
galaxies surrounding the Milky Way today. The metallicities of S1 and
S2 ([Fe/H] $\approx -1.78$ and $-1.91$ respectively) are consistent
with stellar masses of $\sim 10^{5.5}$ through the mass-metallicity
relation~\citep{Ki13}. Although such masses are slightly lower than
our modelling suggests, it must be remembered that there is
considerable scatter in both the abundance matching and the redshift
dependence of the mass-metallicity relations.

We identified four further pieces of substructure; namely, two moving
groups or clumps (C1 and C2) and two substructures (S3 and S4). The
latter two share some similarities with S1 and are also probably
streams in the later stages of disintegration. As all our
substructures are nearby, the member stars are candidates for high
resolution spectroscopic follow-up to provide abundances and ages. The
larger substructures probably extend beyond the volume accessible in
SDSS-Gaia, and it would be valuable to trace their full extent.

The overall aim of activity in this field is to provide an assessment
of the fractional mass in substructure as a function of Galactic
position and metallicity.  Nevertheless, in Rutherford's words, the
'Stamp Collecting' is still insightful.  It is useful to understand
the largest substructures in the nearby halo and the the nature of the
accretion events that gave rise to them. Our matches with the remnants
of accretion events in libraries of numerically constructed stellar
halos have provided insights, but they are not perfect -- for example,
we failed to reproduce the full broadness of the azimuthal velocity
distribution in the case of S1. In fact, it was difficult
to find perfect matches, even though our task was eased by the absence 
of a Galactic disk in the library of \citet{Am17}. This suggests that 
the problem of matching substructures in Gaia DR2 to accreted subhalos 
in simulations may be challenging.

\section*{acknowledgements}
GCM thanks Boustany Foundation, Cambridge Commonwealth, European \&
International Trust and Isaac Newton Studentship for their support on
his work. NWE thanks the Max Planck Institute for Astrophysics for
hospitality during a working visit. We are grateful to the anonymous
referee helped us improve the presentation and to Amina Helmi for
providing data on halo stream stars. The research leading to these
results has received partial support from the European Research
Council under the European Union's Seventh Framework Programme
(FP/2007-2013) / ERC Grant Agreement no. 308024.


\label{lastpage}
\end{document}